\documentclass[10pt]{IEEEtran}

\usepackage{float}
\usepackage{subfigure}
\usepackage{stfloats}
\usepackage{cite}
\usepackage{amsmath,amssymb,amsfonts}
\usepackage{graphicx}
\usepackage{textcomp}
\usepackage{mathrsfs}
\usepackage{cases} 
\usepackage{algorithm}
\usepackage{algorithmicx}
\usepackage{algpseudocode}
\usepackage{color}
\usepackage{amsmath}

\newcommand{\be}{\begin{equation}}
\newcommand{\ee}{\end{equation}}
\def\bee#1\eee{\begin{align}#1\end{align}}
\newcommand{\bse}{\begin{subequations}}
\newcommand{\ese}{\end{subequations}}

\ifCLASSINFOpdf
\else
\fi

\begin{document}
	
	\title{Technical Report for A Joint User Scheduling and Trajectory Planning Data Collection Strategy for the UAV-assisted WSN}

	\author{Xindi Wang, Chi-Tsun Cheng, Lei Deng, Xiaojing Chen, Fu Xiao}

\maketitle

\begin{abstract}
Unmanned aerial vehicles (UAVs) are usually dispatched as mobile sinks to assist data collection in large-scale wireless sensor networks (WSNs). However, when considering the limitations of UAV's mobility and communication capabilities in a large-scale WSN, some sensor nodes may run out of storage space as they fail to offload their data to the UAV for an extended period of time. To minimize the data loss caused by the above issue, a joint user scheduling and trajectory planning data collection strategy is proposed in this letter, which is formulated as a non-convex optimization problem. The problem is further divided into two sub-problems and solved sequentially. Simulation results show that the proposed strategy is more effective in minimizing data loss rate than other strategies.
\end{abstract}


\IEEEpeerreviewmaketitle

\section{Introduction}
By introducing an unmanned aerial vehicle (UAV) as a mobile sink into large-scale wireless sensor networks (WSNs), efficient and effective data collections can be achieved for sensor nodes (SNs) that are deployed in hard-to-reach non-urban areas. In this way, the UAV can establish direct communication links through the Line of Sight (LoS) air-to-ground channels to the served SNs, and thus the communication quality and the lifetime of the WSN can both be improved. 

UAV-assisted WSNs have attracted growing interest in recent years. The existing works mainly focus on optimizing the UAV trajectory and SNs' running parameters to improve the network performances. Zeng {\em et al.} \cite{Zeng2016}, \cite{Zeng2017} formulate the UAV trajectory planning as a non-convex optimization problem, and solve it by a successive optimization method. Zhan {\em et al.} \cite{Zhan2019, Zhan2018Access, Zhan2018TVT} further incorporate other critical parameters of wireless communications into their optimization process, including the status of the SNs, communication outage probability, error estimation, etc., and propose a variety of UAV path planning strategies accordingly. 
Furthermore, not limited to WSNs, UAVs have been widely applied to various wireless communication scenarios \cite{LiuWCL2019,Chen2020TVT,Wang2020TC}, and yield promising results.

However, considering that a UAV with limited mobility and communication capabilities can only serve a finite number of SNs at a time, user (i.e. SN) scheduling is therefore a critical issue for WSNs consisting of a large number of SNs, which may cause some SNs to lose the data they just sampled due to storage exhaustion \cite{Li2019Jsac} as they cannot offload their data to the UAV for a long time. Such a critical issue has often been ignored in previous studies by assuming either $1$) that the UAV can communicate with all SNs in the scenario all the time \cite{Jiang2019,Zeng2016,Zeng2017,LiuWCL2019,Wang2020TC}; $2$) that the UAV only serve one user (SN) at a time \cite{Zhan2019,Zhan2018Access,Zhan2018TVT,Chen2020TVT}, which are not always valid in real-life applications.

To fill the above research gap, a joint user scheduling and trajectory planning data collection strategy for a UAV-assisted large-scale WSN is proposed in this letter, which is executed in a time-division manner. At each time slot, a UAV will schedule the users (i.e. SNs) to communicate with at the next time slot and determine the next way-point to collect the data from the involved SNs. Meanwhile, the orthogonal frequency division multiple access (OFDMA) technology is also considered in the UAV-SNs communication processes, which enables concurrent multi-SN transmissions through flexible bandwidth allocation.
Specially, different from existing OFDMA-WSN studies \cite{Yang2017GlobalCom,Ren2019} that focus on optimizing the allocation of wireless resource blocks (e.g. communication channels) in a static network scenario, the communication system changes all the time as the UAV moves in this letter, which can be treated as an important complement to existing OFDMA-WSN researches. 

\section{System Model and Problem Formulation}
\subsection{System Model}
Consider an area of interest (AoI) where $|\Omega_{all}|$ SNs are randomly and evenly distributed.
For each SN $i$ in the AoI, its location (i.e. $\textbf{L}_i=[L_x(i),L_y(i)]\in\mathbb{R}^{2\times 1}$) is fixed and has been prestored in the UAV. The data queue length of SN $i$ at time slot $t$, denoted as $B_i(t)$, depends on both the data sampling rate (i.e. $s$) and the data uploading throughput with the UAV (i.e. $Thr_i(t)$), which is given as
\begin{align}
B_i(t)=\min\{B_i(t-1)+s-Thr_i(t),B\}.\label{eq:data_queue}
\end{align}
Notice that if the queue length $B_i(t)$ exceeds the storage capacity $B$, the data sampled later will be discarded due to the storage exhaustion. In addition, it is assumed that the data queue of SN $i$ can be estimated by the UAV based on the known sampling rate and the historical communication records with each SN.

In order to facilitate analysis, the behaviors of the UAV during the entire data collection process is modeled in a time-division manner, so that the trajectory of the UAV can be represented as a serial of way-point coordinates $\{\textbf{L}_u(t)\}_{t=1}^{T}$, of which $\textbf{L}_u(t)=[L_{ux}(t),L_{uy}(t)]$ denotes the way-point coordinate of the UAV projected on the ground at the $t$-th time slot (obtained through the equipped GPS), and the flight duration $T$ of the UAV in a data collection task should meet the following condition, namely
\begin{align}\label{eq:endurance}
T\leq T_e,
\end{align}
where $T_e$ is the maximum flight time of the UAV.
Meanwhile, the displacement of the UAV from the $t-1$-th time slot to the $t$-th time slot is governed by
\begin{align}
d(t)=||\textbf{L}_u(t)-\textbf{L}_u(t-1)||\leq V_{\max}\times\tau,\label{eq:flight_range}
\end{align}
where $V_{\max}$ denotes the maximum speed of the UAV and $\tau$ is the interval of adjacent time slots.

Assume that the UAV is flying at the constant altitude $H$, and the projected distance between the UAV and SN $i$ at time slot $t$ is measured as $d_{u,i}(t)=||\textbf{L}_i-\textbf{L}_u(t)||$. Thus the communication rate of SN $i$ at time slot $t$ can be expressed as 
\begin{align}
r_i(t)=W\log_2(1+\alpha_i\frac{p}{\sigma_i^2 (H^2+d_{u,i}^2(t))}),\label{eq:trans_rate}
\end{align}
where $\alpha_i$ denotes the real-time reference channel power gain at $d_{u,i}=1$m in time slot $t$ and $p$ denotes the transmission power that is configured as the same value initially. In addition, due to the fact that the distance between the SNs communicating with the UAV is relatively close and $\tau$ in \eqref{eq:flight_range} is short enough, the channel status (i.e. $\sigma_i^2$) between each UAV-SN communication link is assumed to be the same and remain unchanged throughout a time slot. Therefore, it is also assumed that any SN $i$ can establish a reliable communication connection with the UAV whenever
\begin{align}
d_{u,i}(t)\leq d_c,\label{eq:communication_range}
\end{align}
where $d_c$ is the average effective communication distance\footnote[2]{A UAV-ground communication link is considered reliable if $r_i(t)\geq \epsilon$, and given the same environment, the rate threshold $\epsilon$ mainly depends on the communication distance.} between SN and UAV projected on the 2D plane.

Meanwhile, by adopting OFDMA technology, the UAV can pre-divide communication resources into $n$ equal and orthogonal resource blocks in advance, of which the bandwidth allocated to each communication SN is $W$ (in \eqref{eq:trans_rate}).
In this way, the UAV using OFDMA technology can support concurrent data transmissions with up to $n$ surrounding SNs that meet the condition in \eqref{eq:communication_range} at each time slot $t$, which is formulated as
\begin{align}
|\Omega_o(t)|\leq n,\ \forall t,\label{eq:set}
\end{align}
where $\Omega_o(t)$ denotes the set of SNs chosen by the UAV to communicate with at time slot $t$. So, the throughput of SN $i$ at time slot $t$ is given by 
\begin{align}\label{eq:Thr}
Thr_i(t)=\min\{r_i(t)\tau_c,B_i(t)\},\ \forall i \in \Omega_o(t),
\end{align}
in which $\tau_c$ denotes the duration of a data communication process within one time slot, which is assumed that $\tau_c\ll \tau$.

\subsection{Problem Formulation}
Under the considered application scenario\footnote[4]{Notice that our work can also extended to the scenario with the probabilistic LoS channel model \cite{You2020,Duo2020}. However, due to the page limit, corresponding details are given in section \uppercase\expandafter{\romannumeral4} of our technical report[].}, each SN $i\in\Omega_{all}$ will first store the sampled data in the storage with data queue length $B_i(t)$, and wait for the UAV to perform the data collection task.
However, when considering the physical limitations of the UAV in maneuvering (i.e. the constraints \eqref{eq:endurance}, \eqref{eq:flight_range}) and communication capacities (i.e. the constraints \eqref{eq:communication_range}, \eqref{eq:set}) in a large-scale AoI, it can be expected that only a group of SNs can  communicate with the UAV at the same time, which may cause some SNs to discard the data just sampled due to exhaustion of storage space \cite{Li2019Jsac} because they cannot communicate with the UAV for a long time.
The amount of data be discarded by SN $i$ at the $t$-th time slot is given by
\begin{align}
D_i(t)=\max\{0,B_i(t-1)+s-B\}.\label{eq:data_loss}
\end{align}

Therefore, our goal is to design a data collection strategy for the UAV-assisted WSN to minimize the data loss rate $\eta$ due to storage exhaustion during the entire data collection task, which is given as
\begin{align}
\textup{P1}:&\min\ \eta= \sum_{t=1}^{T}\sum_{i\in \Omega_{all}}\frac{D_i(t)}{s*T}\nonumber\\
&\textup{s.t.}\ \eqref{eq:data_queue}-\eqref{eq:Thr}.\nonumber
\end{align}

\section{An Efficient Strategy to Solve Problem $\textup{P1}$}
For the sake of non-convexity of the objective function and constraints (\eqref{eq:trans_rate}, \eqref{eq:set}), it is hard to solve problem $\textup{P1}$ using general optimization methods.
In this section, we proposed a joint \textbf{U}ser scheduling and \textbf{T}rajectory planning \textbf{S}trategy (named as UTS) to approximate solve problem $\textup{P1}$ in an efficient way, in which the original problem will be divided into two sub-problems and solved successively.

Assume the location of the UAV at time slot $t-1$ is given as $\textbf{L}_u(t-1)$. To reduce the amount of data loss at time slot $t$ (i.e. $\sum_{i\in \Omega_{all}}D_i(t)$), the most intuitive way is regarding SNs with the smallest sum of available storage space within the current coverage area as the communication objects (i.e. $\Omega_o(t)$) at the next time slot $t$, and then finding the optimal next waypoint (i.e. $\textbf{L}_u(t)$) to 
achieve the maximum data throughput with the SNs in $\Omega_o(t)$ for further releasing their storage space, which can be modeled as sub-problem $\textup{sP1}$ and sub-problem $\textup{sP2}$, respectively.
\subsection{Solve Sub-Problem $\textup{sP1}$}
Searching for a SN set which is within the coverage range (i.e. jointly considering the moving capability \eqref{eq:flight_range} and the reliable communication distance \eqref{eq:communication_range}) of the UAV and owns the smallest sum of available storage space under the constraints of communication resources (i.e. \eqref{eq:set}) at time slot $t$ is the goal of problem $\textup{sP1}$, which is formulated as
\begin{align}
\textup{sP1}:&\Omega_o(t)=\mathop{\arg\min}\sum (B-B_i(t)), \label{eq:problem1}\\
&\textup{s.t.}\ \ \
\eqref{eq:data_queue}, \eqref{eq:flight_range}, \eqref{eq:communication_range}, \eqref{eq:set}. \nonumber 
\end{align}

However, due to the fact that problem $\textup{sP1}$ cannot be solved efficiently using general optimization methods, a heuristic-based method is considered here to obtain a sub-optimal solution with a reasonable computational complexity.
First of all, the location constraints of SNs in $\Omega_o(t)$ (i.e. \eqref{eq:flight_range} and \eqref{eq:communication_range}) are re-expressed as following forms, namely
\begin{numcases}{}
||\textbf{L}_i-\textbf{L}_u(t-1)||\leq V_{\max}\times\tau + d_c,\ \forall i \in \Omega_o(t).\label{eq:next_node}\\
||\textbf{L}_i-\textbf{L}_{u}(t)||\leq d_c,\ \ \ \ \ \ \ \ \ \ \ \ \ \ \ \ \ \ \ \ \ \forall i \in \Omega_o(t).\label{eq:constraintofSN}
\end{numcases}

\begin{figure}[htp]
		\centering
		\includegraphics[width=0.35\textwidth]{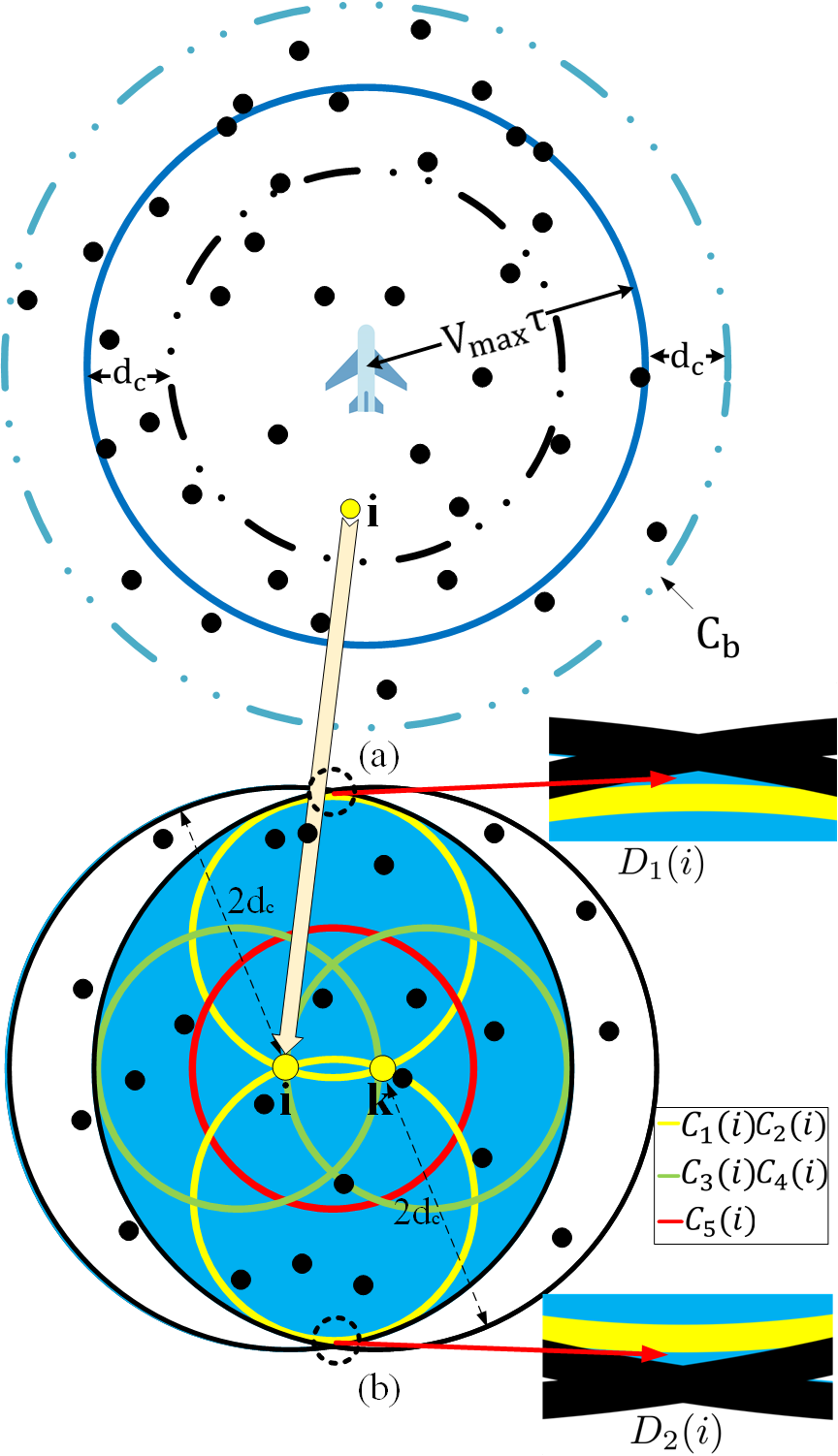}
		\caption{An illustration of (a) the application scenario; (b) the constructed circular areas $\{C_m(i)|m\in \{1,\cdots ,5\}\}$ based on SN $i$ and SN $k$.}
		\label{fig:1}
\end{figure}

Therefore, with a given $\textbf{L}_u(t-1)$, problem \textup{sP1} can be rephrased as the following geometry problem \textup{sP1*}: 

Problem \textup{sP1*}: \textbf{With a given big circular area $C_b$ with $\textbf{L}_u(t-1)$ as its center and $V_{max}\tau+d_c$ as its radius (see \eqref{eq:next_node}), how to get an optimal circular area $C_s$ with $d_c$ as its radius (see \eqref{eq:constraintofSN}) inside $C_b$ which covers at most $n$ SNs (see \eqref{eq:set}) with the smallest sum of available storage space. Then, the set of the involved $n$ SNs are regarded as the $\Omega_o(t)$ that we are searching for.} The graphical illustration is shown in Fig.~\ref{fig:1}(a)\footnote[2]{In order to facilitate the analysis, it is assumed that $V_{max}\tau$ is larger than $d_c$ here. However, even for some special cases where $V_{max}\tau$ is smaller than $d_c$, it is only necessary to update the SN set $\Omega_{d_{max}-dc}$ and its related definition mentioned in Algorithm.~1 as $\Omega_{d_{max}}$: the set of SNs which are within distance $V_{\max}\times\tau$ from $\textbf{L}_u(t-1)$}.

Instead of traversing virtually infinite number of smaller circles with radius $d_c$ inside $C_b$, an efficient search method is proposed, which constructs several alternative circular areas with radius $d_c$ and picks out the one with the best performance. Then, SNs within the chosen optimal circular area are set as a sub-optimal solution of problem \textup{sP1*}, which is graphically represented in Fig.~\ref{fig:1}(b). Meanwhile, the corresponding pseudo code of the proposed search method is given in Algorithm.~\ref{alg:sp1}.

\begin{algorithm}
	\caption{An Efficient Search Strategy}
	\label{alg:sp1}
	\begin{algorithmic}[1] 
		\State \textbf{Initialization:} $\Omega_{opt(i)}=\emptyset$: the constructed SN set with respect to SN $i$;\ \  $\Omega_{d_{max}-dc}$: the set of SNs which are within distance $V_{\max}\times\tau-d_c$ from $\textbf{L}_u(t-1)$.
		\For{each SN $i$ $\textbf{in}$ $\Omega_{d_{max}-dc}$}
		\State Find the partner SN (denoted as SN $k$) for SN $i$, and then put SNs $\{i,k\}\rightarrow$ $\Omega_{opt(i)}$.
		\State Construct the search region $\mathcal{F}(i)$ based on the SN $i$ and SN $k$.
		\State Find the optimal circular area $C^{*}(i)$ with radius as $d_c$ inside $\mathcal{F}(i)$, and put the top $n-2$ SNs with the smallest available storage space from $C^{*}(i)$ into $\Omega_{opt(i)}$.
		\EndFor
		\State \textbf{Output:} Pick out the SN set $\Omega_o(t)$ from $\{\Omega_{opt(i)}|i\in \Omega_{d_{max}-dc}\}$ based on \eqref{eq:problem1}.
	\end{algorithmic}
\end{algorithm}

Further details on steps $3-5$ of Algorithm.~\ref{alg:sp1} are given as follows.

$\blacksquare$ Step $3$:The main idea in this step is to find a SN (denoted as SN $k$) that owns the smallest available storage space and also can communicate with the UAV together with each SN $i\in\Omega_{d_{max}-dc}$, which can be formulated as the following problem, namely
\begin{align}
SN_{k} \leftarrow &\arg\min{(B-B_j(t))},\nonumber\label{eq:SNik}\\
&\textup{s.t.}\ ||\textbf{L}_i-\textbf{L}_j||\leq 2d_c.
\end{align}
	
Note that \eqref{eq:SNik} is a necessary condition that a SN $j$ can communicate with the UAV with the SN $i$ in the time slot $t$, which is adopted here to expand the search range as much as possible. 
	
$\blacksquare$ Step $4$: The main idea in this step is to construct a search region $\mathcal{F}(i)$ that clarifies the range of SNs that can communicate with the UAV together with SN $i$ and SN $k$.
	
At first, two big circular areas are constructed with the coordinates of SN $i$ and SN $k$ (i.e. $\textbf{L}_i$ and $\textbf{L}_k$) as their centers individually and with $2d_c$ as their radii, which are illustrated as the two biggest black circles in Fig.~\ref{fig:1}(b). Then, according to \eqref{eq:SNik}, it is straightforward to prove that SNs located in the overlapping area (denoted as $\mathcal{O}(i)$) of the constructed two circles can communicate with the UAV either with SN $i$ or SN $k$. The overlapping area $\mathcal{O}(i)$ is shown as the blue area in Fig.~\ref{fig:1}(b). 
	
\ \textbf{Proposition 1:} With the given distance between SN $i$ and SN $k$, i.e., $d_{i,k}=||\textbf{L}_i-\textbf{L}_k||$, any SN $j$ attempting to communicate with a UAV simultaneously with SN $i$ and SN $k$ must meet the condition
\begin{align}
\max{\min\{d_{j,i},d_{j,k}\}}\leq \sqrt{\frac{1}{4}d_{i,k}^2+(d_c+\sqrt{d_c^2-\frac{1}{4}d_{i,k}^2})^2},
\end{align}
where $d_{j,i}$ denotes the distance between SN $j$ and SN $i$.
	
\ \ \textbf{Proof:} It is straightforward to prove the above result by the Pythagorean theorem.
	
Here we denote the both end regions of $\mathcal{O}(i)$, i.e., the gaps between $\mathcal{O}(i)$ and the circles of radii $d_c$ passing through SN $i$ and SN $k$ at the same time, as $D_1(i)$ or $D_2(i)$ (shown in Fig.~\ref{fig:1}).
According to Proposition $1$, it is easy to prove that the SNs located in $D_1(i)$ or $D_2(i)$ cannot communicate with UAV together with SNs $i$ and $k$.
Therefore, the search region $\mathcal{F}(i)$ can be obtained by $\mathcal{F}(i)\triangleq\mathcal{O}(i)-D_1(i)-D_2(i)$.
	
$\blacksquare$ Step $5$: The main idea in this step is to obtain an optimal circular area $C^{*}(i)$ inside the search region $\mathcal{F}(i)$, and then put the top $n-2$ SNs with the smallest available storage space from $C^{*}(i)$ into $\Omega_{opt(i)}$.
In order to reduce the computational complexity and cover the area of $\mathcal{F}(i)$ as much as possible, here we construct five alternative circular areas, i.e., $\{C_m(i)|m\in \{1,\cdots ,5\}\}$ with radius $d_c$ inside the region $\mathcal{F}(i)$, and select the one with the best performance.
Among $\{C_m(i)|m\in \{1,\cdots ,5\}\}$, $C_1(i)$ and $C_2(i)$ denote the circular areas with distinct centers while touching both SN $i$ and SN $k$ simultaneously. Then, $C_3(i)$ (\text{or}  $C_4(i))$ is denoted as a circular area that touch SN $k$ (or $i$) and with its corresponding center being collinear with SN $i$ (\text{or} $k$). $C_5(i)$ is the circular area with its center located at the midpoint of SNs $i$ and $k$. 
Based on $\textbf{L}_i$ and $\textbf{L}(k)$, the center coordinates of areas $C_1(i)$ and $C_2(i)$ are respectively derived as
$$ \left\{
\begin{aligned}
L_x(i,k) & = & \frac{L_x(i)+L_x(k)}{2} \mp V\frac{L_y(i)-L_y(k)}{d_{i,k}}, \\
L_y(i,k) & = & \frac{L_y(i)+L_y(k)}{2} \pm V\frac{L_x(i)-L_x(k)}{d_{i,k}},
\end{aligned}
\right.
$$
in which $V=\sqrt{|d_c^2-\frac{1}{4}d_{i,k}^2|}$, and the center coordinates of areas $C_2(i)$ and $C_3(i)$ are respectively derived as
$$ \left\{
\begin{aligned}
L_x(i,k) & = & \textbf{\underline{M}}(L_x(i),L_x(k)) \mp d_c*\frac{|L_x(i)-L_x(k)|}{d_{i,k}}, \\
L_y(i,k) & = & \textbf{\underline{M}}(L_y(i),L_y(k)) \pm d_c*\frac{|L_y(i)-L_y(k)|}{d_{i,k}}, 
\end{aligned}
\right.
$$
in which $\textbf{\underline{M}}(\ )\in \{max(\ ),min(\ )\}$. Note that the selection of the above operators (i.e. $\mp$ and functions $\{max(\ ),min(\ )\}$) are determined by the relative positions between SN $i$ and SN $k$, which can be obtained by simple geometry calculations. Meanwhile, note that if the distance between SNs $i$ and $k$ is equal to $2d_c$, then the aforementioned area, i.e.  $\{C_m(i)|m\in \{1,\cdots ,5\}\}$ will be merged into one. 

Finally, set the circular area that has the smallest sum of available storage space from its top $n-2$ SNs from $\{C_m(i)|m\in \{1,\cdots ,5\}\}$ as $C^{*}(i)$.

$\bigstar$ Computational complexity analysis: step $2$ requires $O(N_R)$ to traverse all SNs in $\Omega_{d_{max}-dc}$, of which $N_R$ represents the average number of SNs in a circular area with a radius of $V_{\max}\times\tau-d_c$. Step $3$ requires $O(N_r)$ to search for the partner SN, in which $N_r$ represents the average number of SNs in a circular area with a radius of $d_c$. 
Step $4$ only involves simple calculations.
Step $5$ requires $O(N_r^2)$ to sort the SNs in each generated circular area $C_m(i), m\in \{1,\cdots ,5\}$, and requires $O(5)$ to pick the one with the best performance. Finally, step $7$ requires $O(N_R)$ to obtain the optimal SN set from $\{\Omega_{opt(i)}|i\in \Omega_{d_{max}-dc}\}$ by comparing the sum of available storage space.

In summary, the computational complexity of the proposed search method is $O(N_R\times(N_r+5*N_r^2+5)+N_R)$, which is considered as practical and highly desirable for the application scenarios under studied in this letter.
\subsection{Solve Sub-Problem $\textup{sP2}$}
After obtaining the SN set $\Omega_o(t)$ by executing Algorithm $1$, sub-problem $\textup{sP2}$ is to get the optimal way-point for the UAV at the $t$-th time slot with the given location $\textbf{L}_u(t-1)$, so that the maximum data throughput between the UAV and SNs in $\Omega_o(t)$ can be obtained. The problem $\textup{sP2}$ is formulated as
\begin{align}\label{eq:sp2}
\textup{sP2}:&\mathop{\max}_{\textbf{L}_u(t)}\ \sum_{i\in \Omega_o(t)}r_i(t)\tau_c\nonumber \\
&\textup{s.t.}\ \ \
\eqref{eq:flight_range}-\eqref{eq:trans_rate}, \eqref{eq:Thr}.
\end{align}

By substituting \eqref{eq:trans_rate} into the above problem, the objective function can be expressed as 
\begin{align}
&\mathop{\max}_{\textbf{L}_u(t)}\ \sum_{i\in \Omega_o(t)}W\log_2(1+\alpha_i\frac{p}{\sigma_i^2 (H^2+d_{u,i}^2(t))})\tau_c\nonumber\\
\Rightarrow	 &\mathop{\max}_{\textbf{L}_u(t)}\ \sum_{i\in \Omega_o(t)}W\log_2(1+\alpha_i\frac{p}{\sigma_i^2 (H^2+||\textbf{L}_i-\textbf{L}_u(t)||^2)})\tau_c.\label{eq:trans_sp2}
\end{align}

Obviously, function \eqref{eq:trans_sp2} is convex with respect to $d_{i,u}^2(t)$, instead of the next way-point of the UAV (i.e. $\textbf{L}_u(t)$). Meanwhile, due to the fact that the first order Taylor expansion of a convex function is the global under-estimator at any point, thus we expand $r_i(t)=W\log_2(1+\alpha_i\frac{p}{\sigma_i^2 (H^2+d_{u,i}^2(t))})$ in \eqref{eq:trans_sp2} with an initial point $d_{i,ui}^2(t)=||\textbf{L}_i-\textbf{L}_{ui}(t)||^2$ in terms of the first-order Taylor expansion as
\begin{align}\label{eq:taylor}
r_i(t)=r_i(t,\textbf{L}_{ui}(t))
+r_i'(t,\textbf{L}_{ui}(t))*\Delta d(t)+\varepsilon,
\end{align}
in which 
\begin{align}
r_i(t,\textbf{L}_{ui}(t))=W\log(1+\alpha_i\frac{p}{\sigma_i^2 (H^2+||\textbf{L}_i-\textbf{L}_{ui}(t)||^2)}),\nonumber
\end{align}
\begin{align}
r_i'(t,\textbf{L}_{ui}(t))=-W\frac{\log_2(e)\frac{\sigma_i^2\alpha_i p}{\sigma_i^4(H^2+||\textbf{L}_i-\textbf{L}_{ui}(t)||^2)^2}}{1+\frac{\alpha_i p}{\sigma_i^2 (H^2+||\textbf{L}_i-\textbf{L}_{ui}(t)||^2)}},\nonumber
\end{align}
\begin{align}
\Delta d(t)=||\textbf{L}_i-\textbf{L}_{u}(t)||^2-||\textbf{L}_i-\textbf{L}_{ui}(t)||^2.\nonumber
\end{align}

Here, if we substitute \eqref{eq:taylor}  without item $\varepsilon$ into original problem, $\textup{sP2}$ will be transformed to a convex optimization problem (denoted as $\textup{sP2*}$) with respect to $\textbf{L}_{u}(t)$ due to the fact that other items are all constants with the initial point $d_{i,ui}(t)$. Therefore, problem \textup{sP2*} can be solved by using general convex optimization methods.

Meanwhile, to reduce the impact of ignoring $\varepsilon$ when transforming \textup{sP2} to \textup{sP2*},
an intuitive method is to choose an initial coordinate $\textbf{L}_{ui}(t)$ which is very likely to be close to the optimal coordinate $\textbf{L}_u(t)$, so that the $\varepsilon=\sum_{n=2}^{\infty}\frac{r_i^{n}(t,d_{ui}(t))}{n!}\Delta d(t)^{n}$ can be negligible as $\Delta d(t)$ tends to $0$.
Therefore, considering the scheduled SN set $\Omega_o(t)$ obtained by Algorithm.~1, one can initialize $\textbf{L}_{ui}(t)$ as the mean of the coordinates of SNs in $\Omega_o(t)$, which ensures communication distances from the UAV to all SNs in $\Omega_o(t)$ are balanced and shorter than $d_c$.
{\color{red}
\section{Extension in Probabilistic LoS Channels}
Here we discuss how to further extend this letter to probabilistic LoS channel scenarios. In this case, the waypoint of the UAV in time slot $t$ should be updated to  $\textbf{q}_u(t)=[\textbf{L}_u(t),L_{uz}(t)]^{T}\in\mathbb{R}^{3\times 1}$ in a given 3-D space. Then, as designed in \cite{You2020,Duo2020}, the LoS probability at each time slot $t$ can be modeled as a function of the UAV-SN elevation angle, which can be expressed in the form of 
\begin{align}
	P(c^{L}_{k,t}=1)=\frac{1}{1+\alpha e^{-b(\theta_{k,t}-\alpha)}},
\end{align}
where $\theta_{k,t}$ denotes the angel between the UAV and SN $k$ in time slot $t$, which is given by
\begin{align}
	\theta_{k,t}=\frac{180}{\pi}arctan(\frac{L_{uz}(t)}{||\textbf{L}_i-\textbf{L}_u(t)||}),
\end{align}
where $\alpha$ and $\beta$ are modeling parameters to be specified. Then, the corresponding NLoS probability can be obtained by 
\begin{align}
	P(c^{L}_{k,t}=0)=1-P(c^{L}_{k,t}=1).
\end{align}

In this case, the large-scale channel power gain between the UAV and SN $k$ in the time slot $t$, including both the path loss and shadowing, can be modeled as 
\begin{align}
	h_{k,t}=c^{L}_{k,t}h^{L}_{k,t}+(1-c^{L}_{k,t}) h^{N}_{k,t},
\end{align}
where 
\begin{align}
	h^{L}_{k,t}=\beta_0 d^{-\alpha_L}_{k,t},\ h^{N}_{k,t}=\mu\beta_0 d^{-\alpha_N}_{k,t}
\end{align}
denote the channel power gains in LoS and NLoS cases, respectively. Meanwhile, $d_{k,t}$ can be calculated as
\begin{align}
	d_{k,t}=\sqrt{||\textbf{L}_i-\textbf{L}_u(t)||^2+L_{uz}^2(t)}.
\end{align}

Considering that the goal of this letter is to design a data collection strategy for the UAV-assisted WSN to minimize the data loss rate $\eta$ due to storage exhaustion during the entire data collection task. So, in each time slot $t-1$, the UAV will schedule a set of SNs (denoted as $\Omega_o(t)$) which owns the smallest sum of available storage space under the constraints of communication resources (i.e. \eqref{eq:set}), and then searches for the  next optimal waypoint (i.e. $\textbf{q}_u(t)$) of the UAV to maximize the data transmission throughput based on the probability LoS channel model. Obviously, the former issue can still be solved by the Algorithm 1 we designed, while the latter issue can be modeled as the following optimization problem.
\begin{align}\label{eq:sp2}
\textup{P2}:&\mathop{\max}_{\textbf{q}_u(t)}\ \sum_{i\in \Omega_o(t)}r_i(t)\tau_c,\nonumber \\
&\textup{s.t.}\ \ \
\eqref{eq:flight_range}, \eqref{eq:trans_rate}, \eqref{eq:Thr}, (20),
\end{align}
where
\begin{align}
	r_i(t)=W\log_2(1+\alpha\frac{h_{i,t} p}{\sigma_i^2 \Gamma}).
\end{align}

The problem P2 can be solved by methods proposed in \cite{You2020,Duo2020}.
}

\begin{figure*}[t]
	\centering
	\includegraphics[scale=0.42]{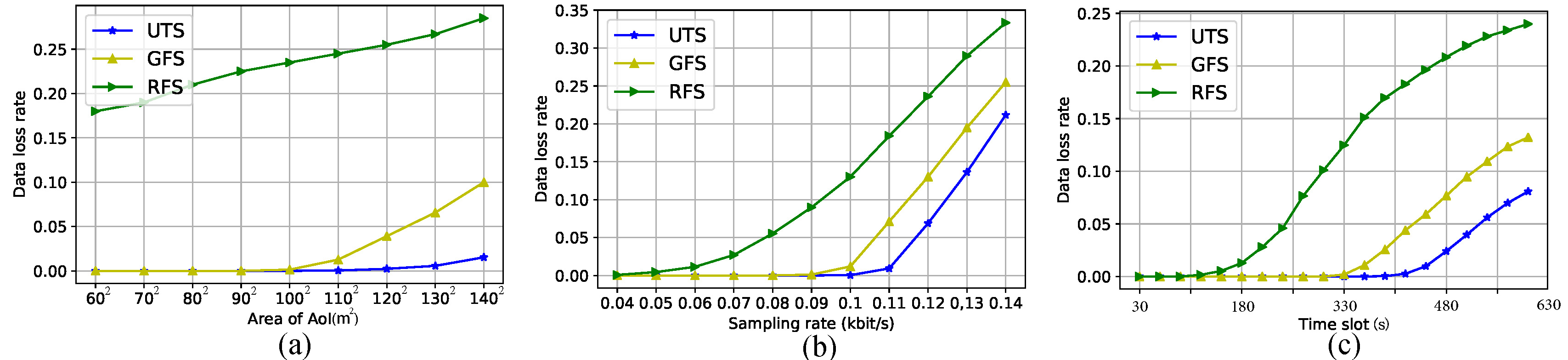}
	\caption{The data loss rate vs. (a) area of AoI with sampling rate $s=100$bit/s over $500$ time slot, (b) sampling rate in a $100\times 100$ $m^2$ scenario over $600$ consecutive time slots, (c) running time with sampling rate $s=120$bit/s in a $100\times100$ $m^2$ scenario from $300$ individual simulations.}
	\label{fig:2}
\end{figure*}
\section{Numerical Results}
A simulated UAV-assisted WSN application scenario has been developed to validate the performances of the proposed strategy \textbf{UTS}. It is assumed that there is only one UAV with a fixed altitude to ground $H=20$m and its maximal speed $V_{max}=30$m/s, and a WSN with $|\Omega_{all}|=200$ SNs being uniformly and randomly distributed in a square AoI. For each SN deployed in AoI, the initial length of the data queue follows a Poisson distribution with $\lambda=10$ kbits, and the storage capacity is $B=30$ kbits. The parameters of the communication model are configured as follows, i.e., bandwidth $W=5$MHz, noise power $\delta^2=-100$dBm, transmission power $p=0.05$W, the reference channel power gain $\alpha_i$ follows a random uniform distribution with range [-55dB, -50dB], the average efficient communication radius $d_c=10$m, the number of wireless resources $n=4$, the duration of each time slot is $\tau=1$s where $\tau_c=0.1$s, and the maximum flight time of the UAV $T_e=600$s.

For benchmarking, a \textbf{R}andom \textbf{F}light \textbf{S}trategy (\textbf{RFS}) and a \textbf{G}reedy \textbf{F}light \textbf{S}trategy (\textbf{GFS}) have been introduced in the simulations, in which RFS chooses the next way-point randomly, and then sets the adjacent SNs with the smallest sum of available storage space as the communication objects. In contrast, GFS chooses the location of the SN with the smallest available storage space within the feasible moving area as its next way-point, and then selects the communication SNs with the same criterion as RFS.
The performance comparisons between the strategies mainly focus on the following three aspects, namely sampling rate (Fig.~\ref{fig:2} (a)), scene scale (Fig.~\ref{fig:2} (b)), and running time (Fig.~\ref{fig:2} (c)). 

As shown in the above simulation results, the proposed UTS shows significant advantages in reducing data loss in multiple scenarios. Compared with RFS without the user scheduling process and GFS with the single-SN greedy scheduling, UTS considers searching for a group of SNs with the greatest overall data loss risk for communication scheduling before performing data collection tasks, which is especially important for a large-scale WSN.
In addition, UTS also optimizes each waypoint in the trajectory to increase the data throughput with the scheduled SNs, which leads to the throughput of the UTS at each waypoint being nearly $3$ times of those in the other two strategies. In this way, more storage space of involved SNs is released and the efficiency of data collection mission can also be guaranteed.

\section{Conclusion}
In this letter, a UAV-assisted data collection strategy has been proposed, which takes into account the user scheduling and trajectory planning of the UAV to reduce the data loss rate in the large-scale WSN. Simulation results have shown that networks with the proposed strategy can yield higher performance than those with other strategies under tests.

\bibliographystyle{IEEEtran}

\end{document}